\documentclass{ieeeaccess}
\usepackage{cite}
\usepackage{amsmath,amssymb,amsfonts}
\usepackage{algorithmic}
\usepackage{graphicx}
\usepackage{textcomp}
\usepackage{hyperref}

\usepackage{bm}
\makeatletter
\AtBeginDocument{\DeclareMathVersion{bold}
\SetSymbolFont{operators}{bold}{T1}{times}{b}{n}
\SetSymbolFont{NewLetters}{bold}{T1}{times}{b}{it}
\SetMathAlphabet{\mathrm}{bold}{T1}{times}{b}{n}
\SetMathAlphabet{\mathit}{bold}{T1}{times}{b}{it}
\SetMathAlphabet{\mathbf}{bold}{T1}{times}{b}{n}
\SetMathAlphabet{\mathtt}{bold}{OT1}{pcr}{b}{n}
\SetSymbolFont{symbols}{bold}{OMS}{cmsy}{b}{n}
\renewcommand\boldmath{\@nomath\boldmath\mathversion{bold}}}
\makeatother

\def\BibTeX{{\rm B\kern-.05em{\sc i\kern-.025em b}\kern-.08em
    T\kern-.1667em\lower.7ex\hbox{E}\kern-.125emX}}

\begin{document}
\history{Date of publication xxxx 00, 0000, date of current version xxxx 00, 0000.}
\doi{10.1109/ACCESS.2025.3635294}

\title{Quantum Mixture-Density Network for Multimodal Probabilistic Prediction}
\author{\uppercase{Jaemin Seo}\authorrefmark{1}}

\address[1]{Department of Physics, Chung-Ang University, Seoul 06974, Republic of Korea (e-mail: jseo@cau.ac.kr)}
\tfootnote{This work was supported by a National Research Foundation of Korea (NRF) grant funded by the Korea government (MSIT) (Grant No. RS-2024-00346024 and RS-2023-00255492).}

\markboth
{Author \headeretal: Preparation of Papers for IEEE TRANSACTIONS and JOURNALS}
{Author \headeretal: Preparation of Papers for IEEE TRANSACTIONS and JOURNALS}

\corresp{Corresponding author: Jaemin Seo (e-mail: jseo@cau.ac.kr).}

\begin{abstract}
Multimodal probability distributions are common in both quantum and classical systems, yet modeling them remains challenging when the number of modes is large or unknown. Classical methods such as mixture-density networks (MDNs) scale poorly, requiring parameter counts that grow quadratically with the number of modes. We introduce a Quantum Mixture-Density Network (Q-MDN) that employs parameterized quantum circuits to efficiently model multimodal distributions. By representing an exponential number of modes with a compact set of qubits and parameters, Q-MDN predicts Gaussian mixture components with high resolution. We evaluate Q-MDN on two benchmark tasks: the quantum double-slit experiment and chaotic logistic bifurcation. In both cases, Q-MDN outperforms classical MDNs in mode separability and prediction sharpness under equal parameter budgets. Our results demonstrate an efficiency in probabilistic regression and highlight the potential of quantum machine learning in capturing complex stochastic behavior.
\end{abstract}


\titlepgskip=-21pt

\maketitle

\section{Introduction}\label{sec1}
\label{sec:introduction}
\PARstart{M}{ultimodal} probabilistic observations are common in physical phenomena, notably within quantum mechanics. For example, nuclear reactions and decay processes often exhibit probabilistic multiple reaction channels; spontaneous emission from an excited atom can occur across multiple photon energy modes; and interference patterns arising from multimodal probability distributions are observed in quantum double-slit experiments. Even classical dynamics exhibits stochastic mode transitions, particularly near critical conditions in nonlinear dynamical systems, such as the randomness observed in fusion plasma phase transitions under identical experimental conditions \cite{prr2020_plasma_random, kim2024_plasma_random}. Stock markets also exhibit probabilistic dynamics due to multimodal crowds reacting differently to identical conditions \cite{CHIARELLA2006_stock_crowd, MF2023_stock_crowd, LAM2025_stock_mdn}. Predicting these probabilistic states and distributions in systems that produce different stochastic outputs under identical conditions is crucial. However, analytical probability calculations often become impractical in noisy or complex systems.

\Figure[t!](topskip=0pt, botskip=0pt, midskip=0pt)[width=\linewidth]{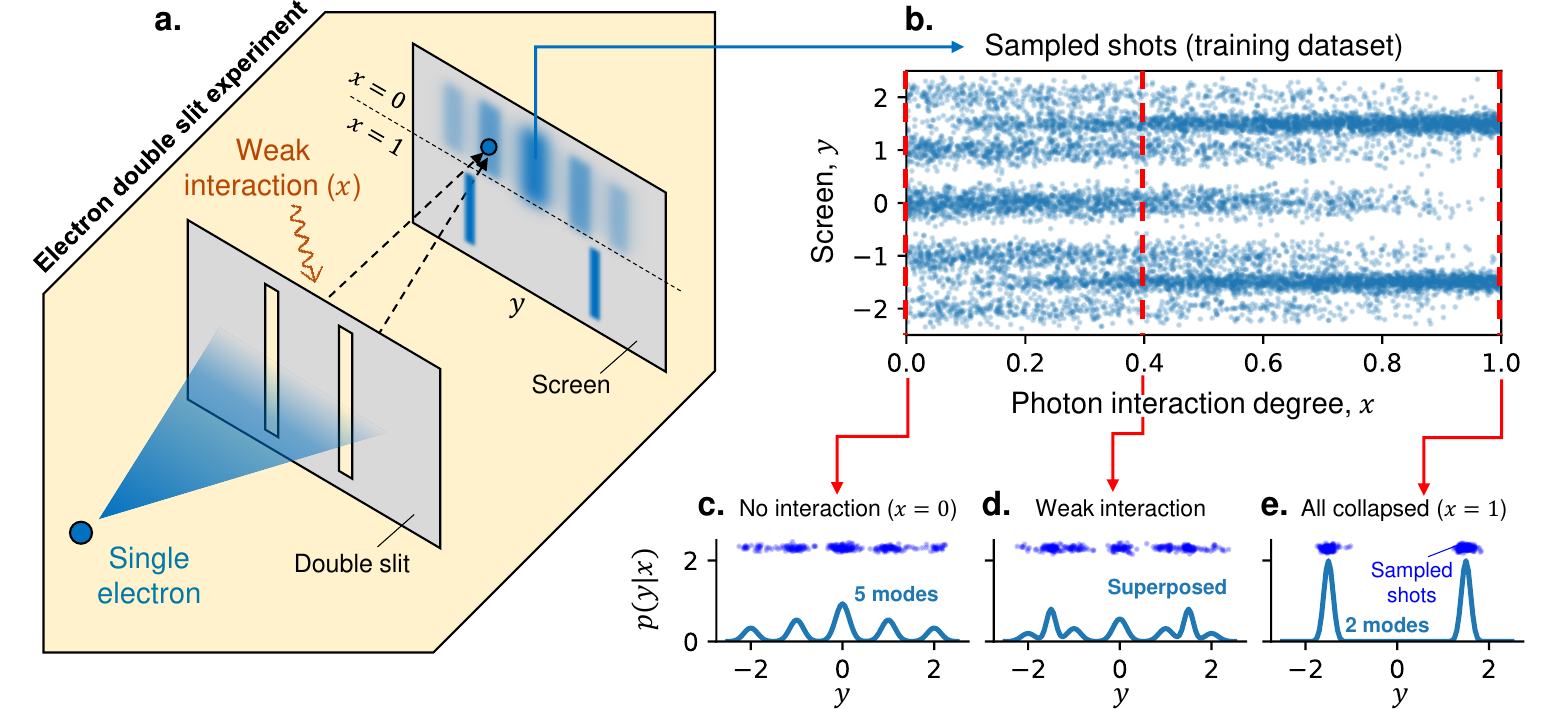}
{ \textbf{Electron double slit experiment and the data distribution.} \textbf{a}, Schematic view of the electron double-slit interference experiment. \textbf{b}, Data distribution at the screen position $y$, depending on the photon interaction degree $x$. \textbf{c-e}, Probability density $p(y|x)$ for where an electron reaches the screen, when $x=0$ (\textbf{c}), $x=0.4$ (\textbf{d}), and $x=1$ (\textbf{e}).\label{fig1}}

Recently, advances in machine learning techniques have facilitated \textit{data-driven} rather than analytic predictions. Data-driven deep learning has successfully predicted states and spectra in nuclear reactions \cite{Seo_NF2021, He2023_nuclear_ml} and the behavior of classical systems near critical conditions \cite{Vega2022, Seo_nature2024}. Nevertheless, it remains challenging to predict outputs for systems capable of producing an unknown number of multimodal states under identical input conditions. A representative example is the multimodal state distribution from an electron double-slit experiment (Figure \ref{fig1}a). One single electron passes through a double slit and is taken at a point on the screen, but when multiple electrons are taken, they form a distribution of interference patterns, as shown in Figures \ref{fig1}a and \ref{fig1}c. However, if a measurement using photons is performed immediately after the double slit, the electron’s wavefunction collapses, resulting in a classical two-peak distribution as illustrated in Figure \ref{fig1}e. If the photon energy is low, the interaction with electrons is weakened, leading to a superposition of interference and collapsed patterns, as depicted in Figure \ref{fig1}d. So the probability density and the number of modes significantly vary depending on the degree of photon interaction (input $x$) at the slits, as shown in Figures \ref{fig1}b–e. 
It is important to note that in this experiment, we can not directly measure the probability density function; instead, we only collect stochastic point data of electrons recorded from each individual shot. 
To learn the probabilities of the multimodal systems from the given dataset, the mixture density network (MDN) has been used \cite{astonpr373_mdn_original, ieee2020_mdn_application, LAM2025_stock_mdn}. Despite its successful application, deep MDNs require a parameter count proportional to the quadratic increase in the maximum number of possible modes. Particularly, the complexity and parameter count become practically infeasible for systems with dozens of modes or more.

With recent advances in quantum computing and algorithms, demonstrating substantial speed-ups over classical computing in tasks such as integer factorization and unstructured search \cite{shor_1999_qc, prl_2017_grover_algo, prl_2009_hhl_algo}, quantum machine learning research has also accelerated \cite{Biamonte2017_qml, Cerezo2022_qml}. Quantum neural networks (QNNs), constructed from parameterized quantum circuits (PQC) \cite{Benedetti_2019_pqc, Hubregtsen2021_pqc, ieee2023_pqc}, are known to offer significantly richer Hilbert space representations compared to classical neural networks \cite{PRR_2021_qnn_expressivity, Wen2024_CP_qnn_expressivity, Panadero2024_SR_qnn_expressivity}. In this study, we introduce a quantum-based MDN (Q-MDN), demonstrating the ability to represent an exponential number of modes with fewer qubits and parameters. This method is particularly suitable for systems with unknown or large numbers of modes, such as those depicted in Figure \ref{fig1}.

\section{Classical multimodal probabilistic predictions}\label{sec2}

\Figure[t!](topskip=0pt, botskip=0pt, midskip=0pt)[width=\linewidth]{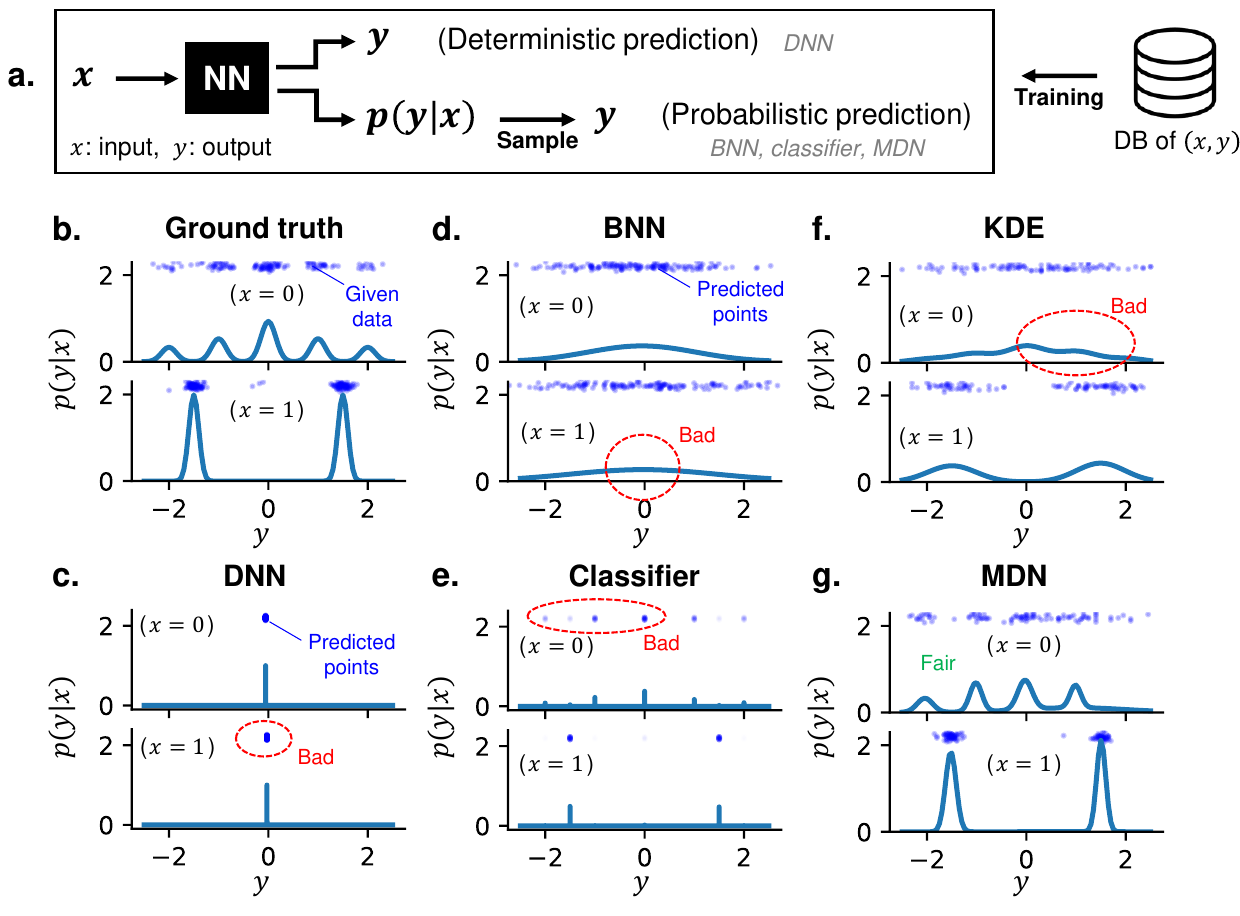}
{ \textbf{Classical approaches to deal with multimodal probabilistic regression.} \textbf{a}, Comparison between the deterministic prediction and the probabilistic prediction. \textbf{b}, Ground truth probability density $p(y|x)$ for the electron double slit experiment, when $x=0$ and $x=1$. \textbf{c-g}, Probability densities estimated by the classical models and the predicted data points, for the deterministic neural network (DNN, \textbf{c}), Bayesian neural network (BNN, \textbf{d}), classifier (\textbf{e}), kernel density estimation (KDE, \textbf{f}), and mixture density network (MDN, \textbf{g}).\label{fig2}}

Classically, data-driven deep learning models have been widely utilized to predict outputs ($y$) for given inputs ($x$). Deterministic neural network (DNN)-based regression models determine a single output for each input, and once trained, the frozen model consistently predicts the same output for identical inputs (refer to Figure \ref{fig2}a). This approach is highly effective for predicting phenomena of deterministic behavior with low uncertainty. However, in systems involving multiple probabilistic modes and broad distributions (such as the double-slit experiments in Figures \ref{fig1} and \ref{fig2}b), different outputs may occur from identical inputs. Training a DNN with such data would result in predicting the weighted average of possible states based on their frequencies, as shown in Figure \ref{fig2}c, which is not our intended goal. Consequently, methods have been developed that predict output distributions rather than single deterministic values. Bayesian neural networks (BNNs), for example, predict not only the output but also its possible deviation to represent uncertainty \cite{Kononenko1989_bnn_original, ieee2022_bnn}. This results in a global scattering of multimodal outputs, as illustrated in Figure \ref{fig2}d. However, Bayesian neural networks cannot distinctly predict multiple modes, particularly struggling with zero-probability regions between modes.

If multimodal prediction is treated as a classification task involving multiple modes, a softmax-based classifier could theoretically predict these modes (Figure \ref{fig2}e). However, this requires prior knowledge of all possible modes and cannot predict continuous mixture distributions, as seen in double-slit experiments. 
Alternatively, a well-established classical method for estimating density from data is kernel density estimation (KDE) \cite{Chen2017_kde}. Compared to other machine learning techniques, KDE offers the advantage of being fast and reproducible when estimating a distribution from a given dataset. However, KDE requires prior knowledge about the bandwidth of the data distribution or relies on bandwidth estimation algorithms, which can significantly alter the resulting density function. In particular, when the data distribution contains multiple peaked modes, KDE tends to estimate a broad and smoothed density function, as shown in Figure \ref{fig2}f. In contrast, mixture density networks (MDNs) can model the conditional probability distribution of the given dataset by predicting the parameters (weights, means, and variances) of a mixture of probability distributions, typically Gaussians $\mathcal{N}_{\mu, \sigma}$ \cite{astonpr373_mdn_original}. Compared to other models, MDNs effectively predict unknown multimodal distributions from given data, as demonstrated in Figure \ref{fig2}g, which will be discussed in detail in the subsequent section.

\section{Classical and quantum mixture-density networks}\label{sec3}

Unlike a deterministic neural network, which produces a single output $y$ for a given input $x$, a mixture density network aims to predict the distribution of possible output values ($p(y|x)$) for each input. Specifically, instead of predicting a single value, an MDN predicts the mean ($\mu_i$), standard deviation ($\sigma_i$), and weight ($\alpha_i$) of each component ($i$) of a mixture probability density function, as shown in Figure \ref{fig3}a. Each pair of $\mu_i$ and $\sigma_i$ forms a Gaussian probability density $\mathcal{N}_{\mu_i, \sigma_i}(y)$, and the distribution function of output values is estimated by the weighted sum of these $n_{modes}$ Gaussian densities (Figure \ref{fig3}c).

\begin{equation}
\begin{split}
    p(y|x) &= \sum_i \alpha_i \mathcal{N}_{\mu_i, \sigma_i}(y) \\
    &= \left. \sum_i \alpha_i(x) \frac{1}{\sqrt{2\pi \sigma_i(x)^2}} \exp \left( -\frac{(y-\mu_i(x))^2}{2 \sigma_i(x)^2} \right) \right|_{\theta_{\alpha, \mu, \sigma}}
\label{eq1} \\
\end{split}
\end{equation}

\Figure[t!](topskip=0pt, botskip=0pt, midskip=0pt)[width=\linewidth]{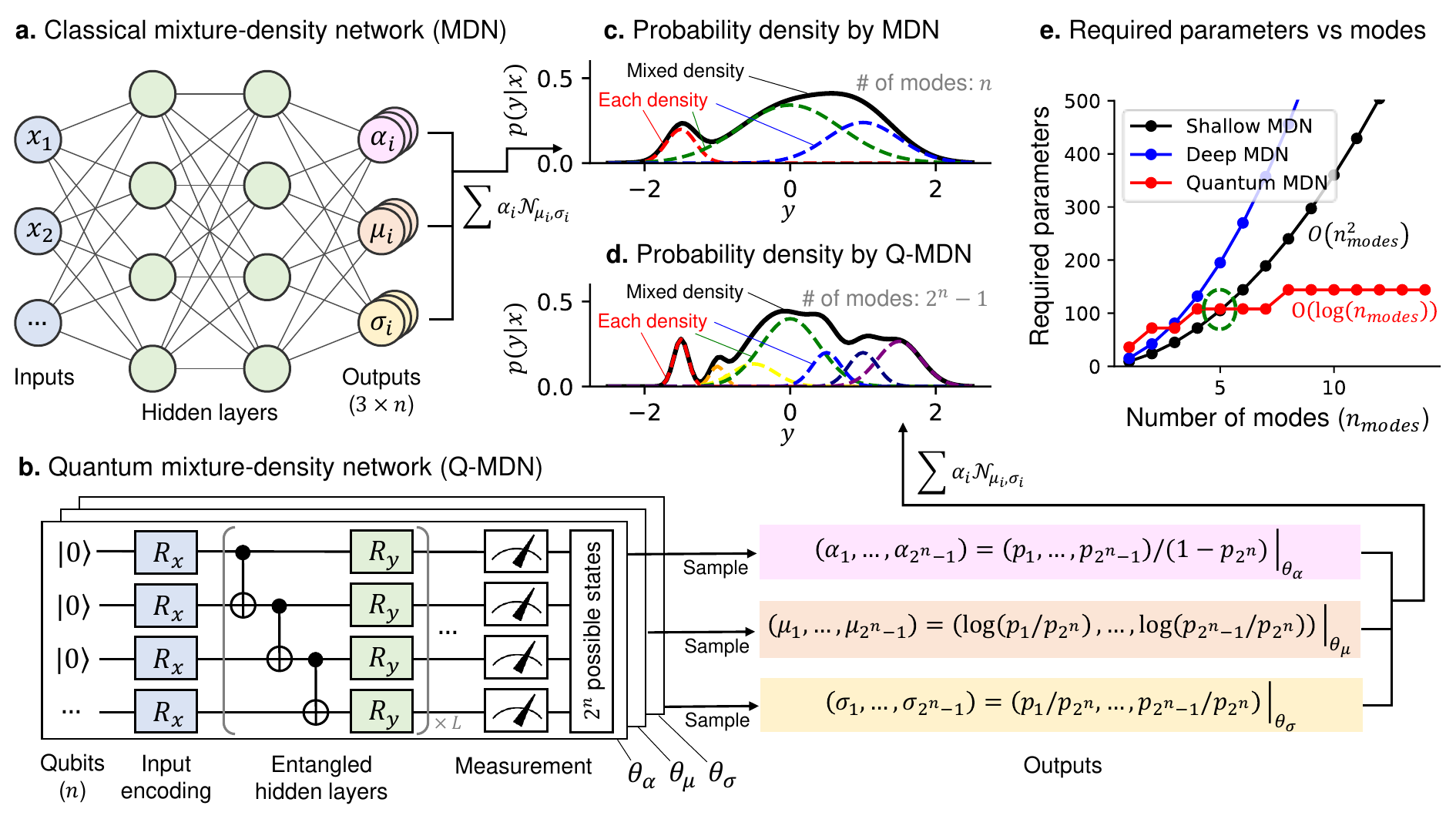}
{ \textbf{Comparison of structural differences between classical MDN and Q-MDN.} \textbf{a}, Architecture of the classical MDN, predicting the weights $\alpha_i$, means $\mu_i$, and standard deviations $\sigma_i$ of a Gaussian mixture. \textbf{b}, Architecture of the Q-MDN and the conversion from qubit states to Gaussian mixture components. \textbf{c}, Constructing probability density through a Gaussian mixture in classical MDN. \textbf{d}, Constructing probability density through a Gaussian mixture in Q-MDN, showing a better mixture expressivity. \textbf{e}, Comparison of the number of parameters required for a given number of modes, between classical MDN (black and blue) and Q-MDN (red). Classical MDN requires parameter counts that grow quadratically with the number of modes, but Q-MDN parameters scale logarithmically with the number of modes.\label{fig3}}

This approach effectively predicts or generates data in cases exhibiting multimodal probabilistic states. The model is trained to ensure that its output distribution accurately explains the given dataset by minimizing a negative log-likelihood loss shown in Equation \ref{eq2}, where $N$ is the batch size.

\begin{equation}
    \mathcal{L} = -\frac{1}{N}\sum_{j=1}^N \log \left[ \sum_i \alpha_i \frac{1}{\sqrt{2\pi \sigma_i^2}} \exp \left( -\frac{(y-\mu_i)^2}{2 \sigma_i^2} \right) \right] 
\label{eq2} \\
\end{equation}

In classical MDNs, the number of peaks in the predicted output distribution is determined by the number of nodes in the output layer. Thus, the output layer must scale proportionally with the maximum number of possible modes in the multimodal system ($3\times n_{modes}$). In addition, the hidden layer size must also scale proportionally to handle this increased complexity. This requirement leads to a quadratic increase in the number of parameters relative to the number of possible modes (Figure \ref{fig3}e), making it impractical for systems with unknown or dozens of modes. While typical MDNs are capable of three or four modes at most, in many-particle quantum systems the number of possible modes increases exponentially, which requires a different approach.

Recently, neural networks based on parameterized quantum circuits (PQCs or QNNs), which utilize unitary transformations of quantum states at each layer, have been employed across various fields \cite{electronics2022_pqc_appl, Ding2024_SR_pqc_appl, SINGH2024_pqc_appl}. QNNs can represent significantly richer Hilbert space representations due to the superposition of multiple quantum states. For instance, while classical NNs (Figure \ref{fig3}a) can only represent feature values equal to the dimension of their output layer, QNNs ($\theta_{\alpha, \text{ }\mu, \text{ or }\sigma}$ in Figure \ref{fig3}b) can provide a much wider range of feature values through an exponential number of possible states determined by the output dimension (or number of qubits). This capability is particularly beneficial for multimodal mixture-density representation tasks, which are of interest in this study. A QNN with $n$ qubits can represent $2^n$ possible states, and the probability ($p_i$) of each state $i$ can be estimated through repeated measurements. However, since these probabilities are constrained between 0 and 1, the last state's probability ($p_{2^n}$) is sacrificed to transform the remaining $2^n-1$ states into logits to represent arbitrary ranges of $\mu_i \in (-\infty, \infty)$ and $\sigma_i \in (0, \infty)$ as follows.

\begin{equation}
    \mu_i = \log \left.\frac{p_i}{p_{2^n}}\right|_{\theta_\mu} \text{, } \sigma_i = \left.\frac{p_i}{p_{2^n}}\right|_{\theta_\sigma}
\label{eq3} \\
\end{equation}

These log-ratios of the basis probabilities enable the effective multivariate regression for predicting multimodal densities \cite{seo2025multivariatequantumregression}. The weights ($\alpha_i$) corresponding to each density are also rescaled so that their total sums to 1 ($\sum_{i=1}^{2^n-1} \alpha_i=1$ and $\alpha_i \in (0, 1)$).

\begin{equation}
    \alpha_i = \left. \frac{p_i}{1-p_{2^n}} \right|_{\theta_\alpha}
\label{eq4} \\
\end{equation}

Each quantum state ($i=1, 2, ..., 2^n-1$) corresponds to a mode of the multimodal mixture density. Three separate QNNs ($\theta_{\alpha, \text{ }\mu, \text{ or }\sigma}$) are required to obtain $\alpha_i$, $\mu_i$, and $\sigma_i$, respectively, as illustrated in Figure \ref{fig3}b. The total mixture probability density is then calculated by the weighted sum of Gaussians for each mode, as depicted in Figure \ref{fig3}d. Compared to classical NNs (Figure \ref{fig3}c), QNNs can represent an exponentially larger number of modes relative to network dimensions, making them capable of covering significantly more or higher-resolution modes. This is advantageous for setting adequate margins in scenarios where the number of modes and their structures are unknown.

The number of parameters in classical MDNs scales quadratically with the dimension of each layer (or the number of distinguishable modes), increasing further with deeper networks, as shown in Figure \ref{fig3}e. In contrast, the parameter count for Q-MDNs is primarily determined by the number of qubits, scaling logarithmically with the number of distinguishable modes. Therefore, Q-MDNs require far fewer parameters than classical MDNs. Since the parameter count directly impacts the computational operations and memory required, Q-MDNs offer superior efficiency in training time and computational resources or enable greater multimodal capability and mode distinction under limited parameter conditions.

\section{Quantum model training for multimodal density prediction}\label{sec4}

\Figure[t!](topskip=0pt, botskip=0pt, midskip=0pt)[width=\linewidth]{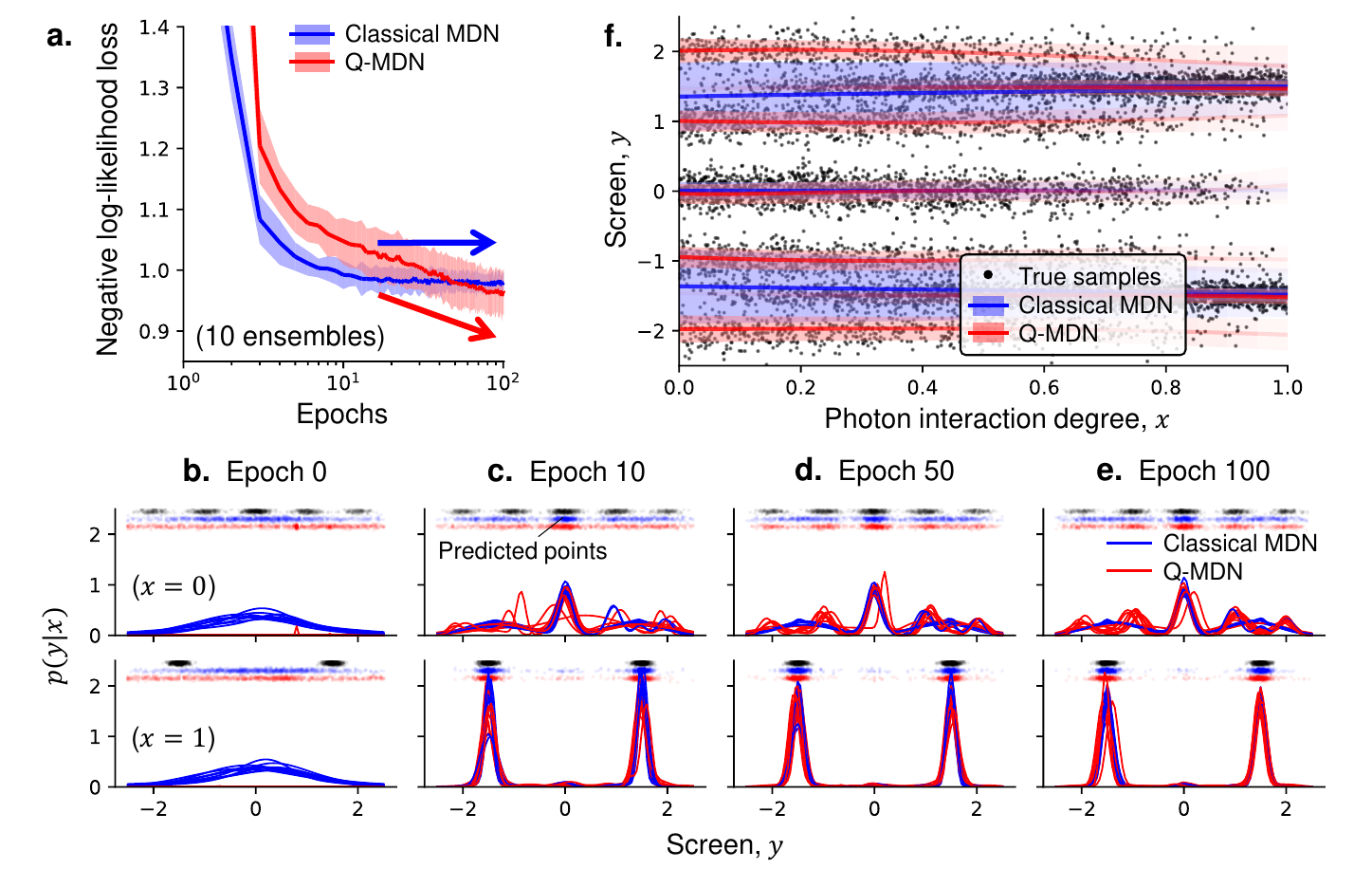}
{ \textbf{Comparison of the performance of classical MDN and Q-MDN on electron double-slit data.} \textbf{a}, Loss histories by epochs for classical MDNs and Q-MDNs, with ten ensemble models each. Classical MDNs show faster reduction and convergence of loss, but Q-MDNs show lower loss in the longer epochs even though it is slower. \textbf{b-e}, The predicted probability densities and sampled data points at each checkpoint; Epoch 0 (\textbf{b}), Epoch 10 (\textbf{c}), Epoch 50 (\textbf{d}), and Epoch 100 (\textbf{e}). Initially, Q-MDN (red) predicts more irregular probability densities than classical MDN (blue), but as the epoch progresses, it distinguishes multi-modes of the ground truth well. \textbf{f}, Comparison of predicted data points of trained MDN (blue) and Q-MDN (red), with ground truth data (black).\label{fig4}}

In this section, we quantitatively analyze the performance of Q-MDNs compared to classical MDNs using the electron double-slit data depicted in Figure \ref{fig1}. To ensure a fair comparison, the number of parameters for both types of MDNs was standardized to approximately 100, as indicated by the green dashed circle in Figure \ref{fig3}e.

Figure \ref{fig4}a shows the loss history during training for classical MDNs and Q-MDNs. The classical MDN exhibits a faster initial reduction in loss, suggesting higher training efficiency in the early stages. However, while classical MDN's loss value converges to a certain level after dozens of epochs, Q-MDN's loss value decreases slowly but continuously, eventually reaching a lower loss than classical MDN. The lower loss observed in the Q-MDN may suggest a greater concern for overfitting on a noisy database; however, similar to the deep double-descent effect, it can also lead to more favorable long-term learning behavior. Figures \ref{fig4}b–e display the estimated $p(y|x)$ distributions and corresponding $y$ samplings at different training epochs, which show differences in the representability of the classical MDNs and Q-MDNs. At around 10 epochs, Q-MDN shows more irregular mode prediction at $x=0$ where there are many modes, but at 100 epochs, Q-MDN shows more detailed mode distinction, which exactly distinguish the 5 modes of the ground truth shown in Figure \ref{fig2}b. Due to its exponentially richer mode representation capability by the entanglement of qubits, the Q-MDN progressively differentiates the correct modes over epochs. Conversely, despite being structured to distinguish up to five modes, the classical MDN with the same number of parameters, differentiates only three modes at $x=0$.

Figure \ref{fig4}f illustrates a sample batch from the true dataset of $(x, y)$ (black dots) alongside predictions generated by the trained classical MDN (blue) and Q-MDN (red). The classical MDN struggles to distinctly separate the multimodal distribution, often merging smaller distribution modes near $x\approx 0$ and $y\approx \pm 1.5$. In contrast, the Q-MDN successfully differentiates even the modes present in distributions with fewer data points.

\section{Application to chaotic bifurcation}\label{sec5}

To further discuss the capability of the Q-MDN, we explore a more nonlinear and complicated example, a dataset of the chaotic bifurcation observed in the logistic map. In the logistic map, as the system parameter (input $x$) varies, a single-amplitude (output $y$) regime bifurcates into multiple amplitude modes, eventually entering a chaotic regime, as illustrated with black dots in Figure \ref{fig5}a. 


\Figure[t!](topskip=0pt, botskip=0pt, midskip=0pt)[width=\linewidth]{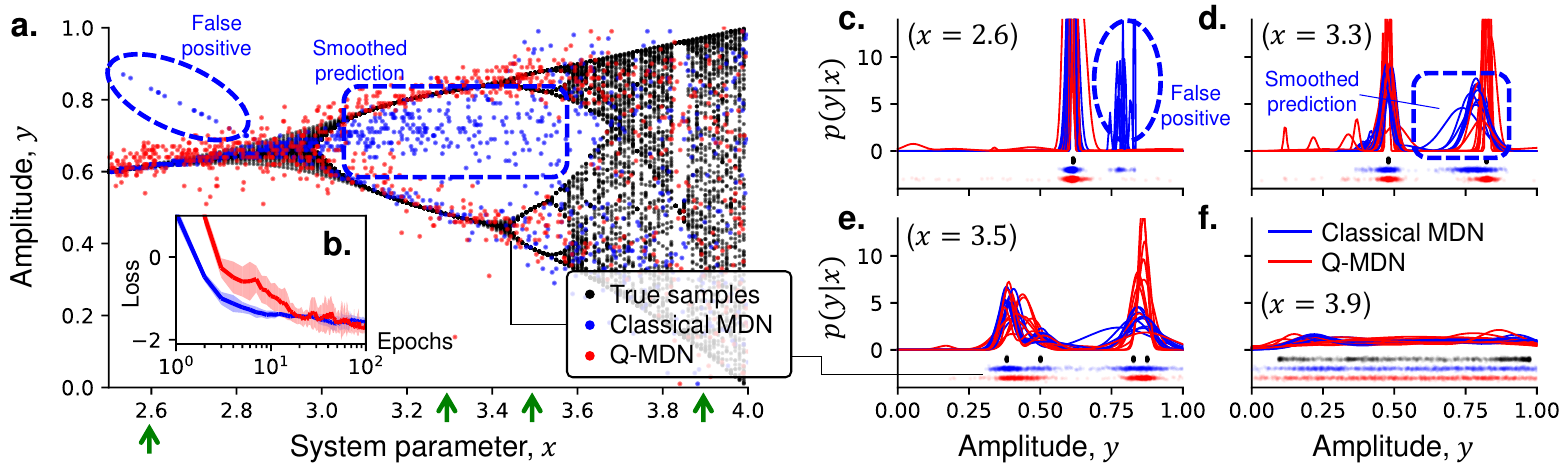}
{ \textbf{Prediction of the chaotic logistic bifurcation using classical MDN and Q-MDN.} \textbf{a}, Along with the ground truth data (black) of the logistic map, data points predicted by classical MDN (blue) and Q-MDN (red). \textbf{b}, Negative log-likelihood loss histories of classical MDNs and Q-MDNs, with ten ensemble models each. \textbf{c-f}, Predicted probability densities by classical MDN (blue) and Q-MDN (red) ensembles, for $x=2.6$ (\textbf{c}), $x=3.3$ (\textbf{d}), $x=3.5$ (\textbf{e}), and $x=3.9$ (\textbf{f}). The ground truth data and the predicted data points are also shown at the bottom of each graph.\label{fig5}}

A distinctive feature of logistic map data is that the number of possible modes can vary significantly depending on the system parameter $x$, ranging from a single mode ($x \approx 2.5$) to chaotic regimes with a continuous (effectively infinite) number of modes ($x \approx 4$). Additionally, the width (broadness) of each mode can also vary significantly, from very sharp modes to broadly spread continuous distributions, presenting unique challenges compared to typical mixture density problems. Here, again, we compare the performance of the classical MDN and Q-MDN with the same number of their parameters.

As can be seen in the loss history in Figure \ref{fig5}b, classical MDN shows fast loss reduction in the beginning, but in the long run, Q-MDN can reach a lower loss. After 100 epochs of training, for each of the 10 ensemble models, classical MDN has an average loss of -1.576, while Q-MDN shows better performance with -1.691. Figures \ref{fig5}c–f show the probability densities $p(y|x)$ predicted by the trained classical MDN and Q-MDN for various system parameters $x$, with ten ensemble models each. At the bottom of each graph, true samples for the given $x$ (black dots) along with points sampled from the predicted probability densities (blue and red dots) are marked. There are two statistically clear differences between the two approaches across these ensembles.

First, for the single-mode case when $x<2.8$, the Q-MDN consistently predicts sharply peaked densities around the correct mode, whereas the classical MDN often yields false-positive predictions at incorrect locations (blue dashed circles in Figures \ref{fig5}a and \ref{fig5}c). This issue arises because 
the classical MDN maintains non-vanishing distributions from subsequently bifurcated modes (at $x \approx 3.3$) even in single-mode regions (at $x \approx 2.5$). In contrast, due to its richer representational capability at the same number of parameters, the Q-MDN more effectively captures the sudden changes in probability densities before and after bifurcation.

\Figure[t!](topskip=0pt, botskip=0pt, midskip=0pt)[width=\linewidth]{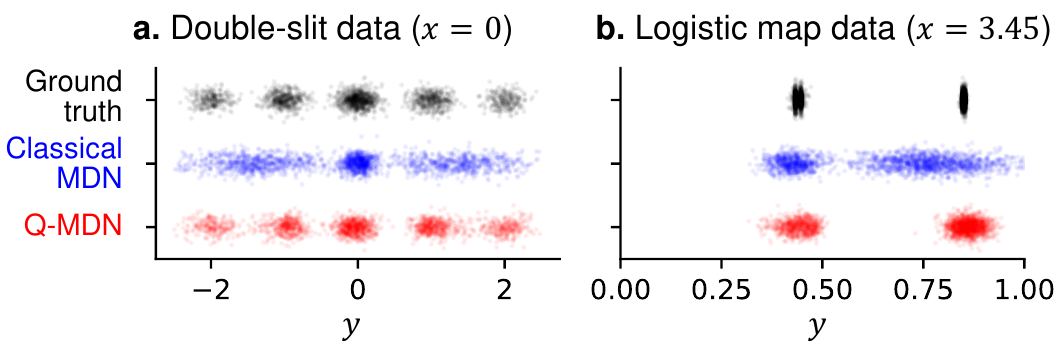}
{ \textbf{Comparison of the resolution capability between the classical MDN and the Q-MDN.} \textbf{a}, Along with the ground truth data (black) of the double-slit experiment data, data points predicted by classical MDN (blue) and Q-MDN (red). \textbf{b}, Along with the ground truth data (black) of the logistic map data, data points predicted by classical MDN (blue) and Q-MDN (red). With the same number of parameters, Q-MDN shows a better separability than classical MDN.\label{fig6}}

Second, the Q-MDN significantly outperforms the classical MDN in capturing sharply peaked, nearly deterministic modes. For instance, at $x=3.3$, the logistic map exhibits two sharply peaked modes. However, the classical MDN predicts a significantly smoothed density, leading to erroneous predictions in the empty region between the modes (blue dashed rectangles in Figures \ref{fig5}a and \ref{fig5}d). On the other hand, the Q-MDN predicts much sharper probability densities, accurately matching the two correct modes. This accuracy arises because the Q-MDN predicts $\sigma_i$ values based on qubit-state probabilities (Equation \ref{eq3}), enabling it to naturally estimate $\sigma_i$ values close to zero. In contrast, the classical MDN employs exponential activations for $\sigma_i$, which require sufficiently large absolute values of logits to yield near-zero predictions. Consequently, as seen in Figure \ref{fig6}a and \ref{fig6}b, when two different modes are closely positioned, the classical MDN often fails to distinguish between them, instead predicting a smoothed, merged density.

\section{Summary and discussion}\label{sec6}

This study demonstrates that quantum neural networks can outperform classical neural networks in regression tasks involving multimodal probability densities, such as those encountered in quantum systems. Classical mixture-density networks (MDNs), traditionally employed for multimodal probabilistic regression, become inefficient when applied to many-body quantum systems with exponentially large numbers of modes. In contrast, quantum mixture-density networks (Q-MDN), based on parameterized quantum circuits, exhibit exponential growth in the number of representable modes with the number of qubits, allowing coverage of numerous modes with fewer qubits and parameters. This approach is especially effective in situations with multidimensional modes or an unknown number of possible modes.

In this study, we applied Q-MDN to the multimodal probabilistic prediction tasks: the electron double-slit experiment and the chaotic bifurcation. Under equal parameter counts, Q-MDN successfully distinguished a greater number of modes compared to classical MDN, demonstrating the effectiveness of quantum approaches for quantum-related problems. However, there remain aspects of this quantum version of the MDN that require further analysis. For example, whether the exponential scaling achieved by increasing the number of qubits remains practically valid in a real noisy quantum device, and how the computational cost is affected by the large number of measurements required to estimate exponentially many multimodal probabilities. Since this work was conducted using a noiseless quantum simulator, we were limited in addressing these questions; however, future experiments on real quantum devices may enable a more thorough investigation in the future.

In addition to multimodal regression shown in this study, this technique can also be used in further machine learning tasks such as reinforcement learning or continuous tree search in cases where multiple decisions are possible in a continuous action space, which will be our future work.

\appendices \label{sec7}
\section{\break Parameterized quantum circuits}\label{subsec7.1}

Parameterized quantum circuits (PQCs), also known as variational quantum circuits, are quantum circuits composed of unitary operations with tunable parameters \cite{Benedetti_2019_pqc, Hubregtsen2021_pqc, ieee2023_pqc}. These circuits form the foundation of quantum neural networks (QNNs) and enable hybrid quantum-classical optimization. A PQC consists of alternating layers of single-qubit rotations and entangling gates, with the general structure of Equation \ref{eq7}.

\begin{equation}
    U(\boldsymbol{\theta})=\prod_{l=1}^L \left( U_{ent}^{(l)} \cdot U_{rot}^{(l)}(\boldsymbol{\theta}^{(l)}) \right)
\label{eq7} \\
\end{equation}

Here, $U_{rot}^{(l)}(\boldsymbol{\theta}^{(l)})$ denotes a layer of parameterized single-qubit rotation gates ($R_x, R_y, R_z$), and $U_{ent}^{(l)}$ applies fixed two-qubit entangling CNOT gates across the qubits. The angle parameter vector $\boldsymbol{\theta}^{(l)}$ at each layer $l$ consists of the rotating angles ($\phi_j^{(l)}, \theta_j^{(l)}, \omega_j^{(l)}$), where $j$ is the qubit index.

For example, a parameterized rotation gate on qubit $j$ is given by

\begin{equation} \label{eq8}
\begin{split}
    &R\!\left(\phi_j^{(l)}, \theta_j^{(l)}, \omega_j^{(l)}\right) 
    = R_z(\omega_j^{(l)}) R_y(\theta_j^{(l)}) R_z(\phi_j^{(l)}) \\
    &= 
    \begin{bmatrix}
    e^{-i(\phi_j^{(l)}+\omega_j^{(l)})/2} \cos\!\left(\tfrac{\theta_j^{(l)}}{2}\right) & 
    -e^{i(\phi_j^{(l)}-\omega_j^{(l)})/2} \sin\!\left(\tfrac{\theta_j^{(l)}}{2}\right) \\
    e^{-i(\phi_j^{(l)}-\omega_j^{(l)})/2} \sin\!\left(\tfrac{\theta_j^{(l)}}{2}\right) & 
    e^{i(\phi_j^{(l)}+\omega_j^{(l)})/2} \cos\!\left(\tfrac{\theta_j^{(l)}}{2}\right)
    \end{bmatrix}.
\end{split}
\end{equation}

The parameter vector $\boldsymbol{\theta}$ is optimized via classical gradient descent based on a loss function (Equation \ref{eq2}) defined over measurements from the quantum circuit. The expressivity of PQCs increases with the number of qubits and circuit depth, enabling the representation of highly complex functions in exponentially large Hilbert spaces.

\section{\break Data of double-slit experiment}\label{subsec7.2}

The electron double-slit experiment is a renowned demonstration of the wave nature of electrons and the quantum mechanical collapse induced by measurement. When a single electron is directed toward two slits, as depicted in Figure \ref{fig1}a, the electron exists as a probability-density wavefunction, simultaneously passing through both slits. Due to interference between the probability-density waves emanating from each slit, the electron distribution on the screen exhibits a multimodal interference pattern. Although each electron individually arrives at a single point on the screen, the cumulative distribution of multiple sequentially emitted electrons gradually forms the interference pattern. This is a clear indication that electrons exhibit wave-like behavior rather than purely classical particle characteristics.

Another intriguing aspect is that if the electron is measured immediately after passing through the slits (using photons), the electron’s wavefunction, previously extended across both slits, collapses into a localized state corresponding to just one slit. In this scenario, the electron behaves classically, traversing only one slit and forming a distribution with two distinct peaks on the screen.

Importantly, the act of observation induces a continuous range of effects depending on the photon’s energy. Low-energy photons interact weakly, leaving the electron’s wave-like nature dominant, whereas high-energy photons strongly interact, collapsing the electron’s wavefunction and accentuating its particle nature. Thus, the degree of photon interaction (input $x$) significantly influences both the expectation value and distribution of electron positions (output $y$) on the screen, even yielding multimodal distributions with varying numbers of modes.

To model this scenario in our study, we define the photon-electron interaction degree as an input parameter $x$, where $x=0$ corresponds to the low-interaction limit (interference pattern with five peaks), and $x=1$ represents the high-interaction limit (classical pattern with two peaks). Under these conditions, the conditional probability distribution $p(y|x)$ determining the electron’s position $y$ on the screen is given as follows:

\begin{subequations}
\begin{align}
    p(y|x) &= (1-x)p_\text{interf}(y) + x p_\text{classical}(y) \\
    \nonumber p_\text{interf}(y) &= 0.35 \mathcal{N}_{0,0.15}(y) + 0.2 \mathcal{N}_{\pm 1,0.15}(y) \\ 
    &+ 0.125 \mathcal{N}_{\pm 2,0.15}(y) \\
    p_\text{classical}(y) &= 0.5 \mathcal{N}_{-1,0.1}(y) + 0.5 \mathcal{N}_{+1,0.1}(y)
\end{align}
\label{eq5}
\end{subequations}

For training neural network models, we generated a dataset by randomly sampling 20,000 pairs of $(x, y)$ based on the above probability density functions.

\section{\break Data of chaotic bifurcation}\label{subsec7.3}

The logistic map describes phenomena in chaotic and coupled systems, such as nonlinear optics \cite{PRA_1987_logistic}, fluid dynamics \cite{logistic_dripping}, or biological populations \cite{science_1997_logistic}, where the distribution of sequential frequencies or amplitudes ($y$) undergoes bifurcation. As the system parameter $x$ is varied, a single-frequency behavior bifurcates, eventually transitioning into a chaotic regime. In this study, the sequential amplitude changes are modeled by the following recurrence relation.

\begin{equation}
    y_{n+1} = x y_n (1-y_n)
\label{eq6} 
\end{equation}

The initial value $y_0$ is set to 0.5. For each given parameter $x$, we discard the initial 5 steps of the $y_n$ series, subsequently using the next 100 data points for analysis. The parameter $x$ is varied within the range from 2.5 to 4, yielding a total of 15,000 data points $(x, y)$ used for training.

\section{\break Model architecture and hyperparameters}\label{subsec7.4}

In this study, we set the number of parameters for the classical MDN and the Q-MDN to be approximately equal, to facilitate a fair comparison between the two models under comparable conditions. The classical MDN was implemented as a multi-layer perceptron consisting of one hidden layer with 5 neurons and an output layer containing 5 neurons. Each neuron in the hidden layer utilizes a tanh activation function, resulting in a total of 105 parameters.

For the Q-MDN, we employed parameterized quantum circuits with 3 qubits in this study. The input variable $x$ was embedded into each quantum circuit wire using $R_x$ rotations. The quantum states subsequently undergo rotation and CNOT entanglement operations across 4 layers (Equation \ref{eq7}). This architecture yields a total of 108 parameters, corresponding to the rotation angles in the quantum circuits. Here, to mitigate barren plateaus, often caused in QNNs \cite{McClean2018_barren_plateau}, we restrict circuit depth and initialize parameters near identity ($\sim \mathcal{N}_{0, 0.05}$). Although adaptive optimizers such as Adam do not prevent barren plateaus, they help stabilize training in shallow circuits where informative gradients exist.

The neural network architectures have been constructed with PyTorch framework \cite{paszke2019pytorch}, and the quantum circuits were modeled with PennyLane \cite{bergholm2022pennylane}. Both models were optimized using the Adam optimizer \cite{kingma2017adam} with a learning rate of $5 \times 10^{-3}$. Training was performed with a batch size of 64, over 100 epochs. For statistically reliable comparisons, 10 ensemble models were used in each case. The codes that generate the data and train the models are available at https://github.com/jaem-seo/Q-MDN.

\begin{IEEEbiographynophoto}{Jaemin Seo} received the B.S. and Ph.D. degrees in Nuclear Engineering in 2017 and 2022, respectively, from Seoul National University, Republic of Korea. From 2022 to 2023, he was a Postdoctoral Researcher with Mechanical and Aerospace Engineering at Princeton University, US. Since 2023, he has been an Assistant Professor in the Department of Physics, Chung-Ang University, Republic of Korea. His current research interests include incorporation of machine learning and physics.
\end{IEEEbiographynophoto}

\EOD

\end{document}